\documentclass[amsmath,amssymb,11pt,superscriptaddress,reprint, preprintnumbers,notitlepage,aps,prl,twocolumn]{revtex4-1}
\pdfoutput=1 
\usepackage[utf8]{inputenc}
\usepackage[english]{babel}
\usepackage{amsmath}
\usepackage{graphicx}
\usepackage{dcolumn}
\usepackage{pbox}
\usepackage{amssymb}
\usepackage{epsfig}
\usepackage[dvipsnames]{xcolor}
\usepackage[normalem]{ulem}
\usepackage{slashed}
\usepackage{amssymb}
\usepackage{verbatim} 
\usepackage{ mathrsfs }
\usepackage{color}
\usepackage[font=small]{caption}
\usepackage[font=small]{subcaption}
\usepackage{url}
\definecolor{MyDarkBlue}{rgb}{0.1, 0.1, 0.8}
\definecolor{SBlue}{rgb}{0.2, 0.4, 0.7} 
\definecolor{MyLightBlue}{rgb}{0.22,0.51,0.9}
\definecolor{MyGreen}{rgb}{0.0, 0.5, 0.0}
\definecolor{BrickRed}{rgb}{0.8, 0.25, 0.33}
\RequirePackage{hyperref}
\hypersetup{colorlinks, citecolor=SBlue,linkcolor=MyDarkBlue, urlcolor=PineGreen}

\makeatletter

\makeatletter
\renewcommand\@makecaption[2]{%
  \par
  \vskip\abovecaptionskip
  \begingroup
  
   \small\rmfamily
    \begingroup
     \samepage
     \flushing
     \let\footnote\@footnotemark@gobble
     \@make@capt@title{#1}{#2}\par
    \endgroup
  \endgroup
  \vskip\belowcaptionskip
}
\makeatother

\DeclareUnicodeCharacter{2212}{-}
\begin{document}
\preprint{MS-TP-24-14}
\title{\vspace{1cm}\Large 
Connecting pseudo-Nambu-Goldstone dark matter with pseudo-Dirac neutrinos in a  left-right symmetry model
}

\author{\bf Sumit Biswas}
\email[E-mail:]{sumit.biswas@okstate.edu}
\affiliation{Department of Physics, Oklahoma State University, Stillwater, OK 74078, USA}
\author{\bf Vishnu P.K.}
\email[E-mail:]{vishnu.pk@uni-muenster.de}
\affiliation{Institut f{\"u}r Theoretische Physik, Universit{\"a}t Münster, Wilhelm-Klemm-Stra\ss{}e 9, 48149 M{\"u}nster, Germany}
\author{\bf Anil Thapa}
\email[E-mail:]{wtd8kz@virginia.edu}
\affiliation{Department of Physics, University of Virginia,
Charlottesville, Virginia 22904-4714, USA}

\begin{abstract}
Stringent constraints from the dark matter (DM) direct detection experiments can be naturally evaded for a pseudo-Nambu-Goldstone boson (pNGB) DM. We propose a realization of pNGB DM in the context of a left-right symmetric model, wherein the neutrinos are pseudo-Dirac in nature. The Dirac mass term for neutrinos arises from two-loop quantum corrections, whereas the Majorana mass terms are generated from Planck-induced corrections. This class of model also provides a parity solution to the strong CP problem without the need for an axion. We show an interesting correlation between the lifetime of the DM and the mass-squared differences between active and sterile neutrinos while maintaining a solution to the strong CP problem.
\end{abstract}

\maketitle

\textbf{\emph{Introduction}.--} 
Weakly interacting massive particles (WIMP) remain one of the most attractive theoretical models for DM, in which DM particles were in the thermal bath with the standard model (SM) particles in the early universe. As the universe expanded, the interaction rate decreased and eventually decoupled, fixing DM relic abundance \cite{Planck:2018vyg}. The scattering process between WIMP and the SM particles imposes a stringent upper limit on the DM-nucleon scattering cross-section \cite{PandaX-4T:2021bab,LZ:2022lsv,XENON:2023cxc}, necessitating suppression of DM-SM scattering while maintaining the DM annihilation cross-section. Thus, to account for the null results from the direct detection experiments, various models of pseudo-Nambu-Goldstone boson \cite{Gross:2017dan,Huitu:2018gbc,Abe:2020iph,Okada:2020zxo,Abe:2021byq,Okada:2021qmi,Abe:2022mlc,Liu:2022evb,Otsuka:2022zdy,Abe:2024vxz} have been proposed. Owing to its Goldstone nature, the DM-nucleon scattering cross-section is proportional to the momentum transfer and leads to a significant suppression in the non-relativistic limit.  

In addition to pursuing a DM candidate, understanding the nature of neutrinos, Dirac vs. Majorana remains a fundamental question in particle physics. Neutrino oscillation experiments are unable to distinguish between the two. An intriguing possibility is that neutrino is a pseudo-Dirac particle \cite{Wolfenstein:1981kw,Petcov:1982ya,Valle:1983dk}: fundamentally Majorana fermion with softly broken lepton number, yet behaving like Dirac fermion. Such a scenario requires tiny mass-squared splitting between active and sterile states, which makes the testability achievable only when the neutrino baseline is astrophysical in the distance. Typical examples in the literature where Dirac mass is naturally small with vanishing Majorana masses at the renormalizable level are Dirac seesaw models \cite{Silagadze:1995tr,Joshipura:2013yba,Gu:2006dc,Ma:2014qra,Valle:2016kyz,CentellesChulia:2018bkz,Jana:2019mez} and left-right symmetric models (LRSM) \cite{Babu:1988yq,Davidson:1987mh,Davidson:1987tr,Bolton:2019bou,Babu:2022ikf}. Tiny nonzero Majorana masses in such models may be induced through higher-dimensional Planck-suppressed operators. 

The primary goal of this letter is to develop a class of left-right symmetric theory based on $SU(3)_c \times SU(2)_L \times SU(2)_R \times U(1)_X \times U(1)_{B-L}$ that naturally leads to pseudo-Dirac neutrinos and pGNB DM. Here the global unbroken $(B-L)$ symmetry of the left-right theory \cite{Babu:1988yq} is promoted to gauge symmetry, crucial to realize pNGB DM as well as to control the amount of active-sterile mass splitting. Unlike conventional LRSM \cite{Pati:1974yy,Mohapatra:1974gc,Mohapatra:1974hk,Senjanovic:1975rk}, the masses of the charged fermions in the model stem from a generalized seesaw mechanism \cite{Davidson:1987mh}; thus offering some insights of the mass hierarchies observed among quarks and leptons.  Interestingly, with the minimum fermion content, which is well motivated as it has natural embedding in $SU(5)\times SU(5)$ GUT \cite{Davidson:1987mi,Cho:1993jb,Mohapatra:1996fu,Lee:2016wiy,Emmanuel-Costa:2011hwa,Tavartkiladze:2016imo,Lonsdale:2014wwa,Babu:2023dzz}, Dirac neutrinos get their tiny masses via two-loop radiative corrections.  

Moreover, the model is well-motivated on several grounds. Parity is spontaneously broken symmetry, unlike in the SM. The model provides a parity resolution to the strong CP problem, obviating the necessity for the Peccei-Quinn symmetry and the resulting axion \cite{Babu:1989rb}. The model can also provide resolution to $W$-boson mass anomaly \cite{CDF:2022hxs} and Cabibbo anomaly \cite{Belfatto:2019swo} (for recent work see Ref.~\cite{Dcruz:2022rjg}). 

\medskip
\textbf{\emph{Model description}--} 
The model we considered is based on the left-right symmetric gauge group $SU(3)_C\times SU(2)_L\times SU(2)_R\times U(1)_X \times U(1)_{B-L}$, wherein the SM fermion fields accompanied with right-handed neutrinos form  $SU(2)_{L,R}$ doublets:
\begin{align}
    Q_{L,R} = (u,\ d)^T_{L,R},
    \hspace{5mm}
    \Psi_{L,R} = (\nu,\ e)^T_{L,R}  \, .
\end{align}
The model also includes three families of vector-like quark and lepton fields $F\equiv(U,D,E)$ that transform as singlets under the $SU(2)$ symmetries, which are needed to generate masses and mixing for the SM charged fermions via a generalized seesaw mechanism \cite{Davidson:1987mh}.
The scalar sector consists of the following Higgs fields
\begin{align}
H_{L,R} & = (H^+,\ H^0)^T_{L,R} , 
\hspace{3mm} \Phi_1 , \hspace{3mm} \Phi_2 \, .
\label{eq:Hspec}
\end{align}
Here $H_{L,R}$ ($\Phi_{1,2}$) fields transform as doublets (singlets) under $SU(2)_{L,R}$ symmetry. The $U(1)_{B-L}$ symmetry
is broken by the vacuum expectation value (VEV) $\langle\Phi_{1,2} \rangle = v_{1,2}/\sqrt{2}$, where as the remaining symmetry is broken down to $SU(3)_c \times U(1)_{em}$ by $\langle H_{L,R}^0\rangle = \kappa_{L,R}/\sqrt{2}$. 
These additional singlet fields $\Phi_{1,2}$ are crucial in realizing a coherent framework for pseudo-Dirac neutrinos and provide a viable candidate for pNGB DM. The quantum number of all these particles under the gauge symmetry of the theory is given in Table~\ref{tab:charge}.
\begin{table}[]
    \centering
    \begin{tabular}{|c|c|c|c|c|c|c|}
    \hline
         & $SU(3)_c$ & $SU(2)_L$ & $SU(2)_R$ & $U(1)_X$ & $U(1)_{B-L}$ \\
         \hline
        $Q_L$ & 3 & 2 & 1 & 1/3 & 1/3  \\
        $Q_R$ & 3 & 1 & 2 & 1/3 & 1/3  \\
        $\Psi_L$ & 1 & 2 & 1 & $-1$ & $-1$  \\
        $\Psi_R$ & 1 & 1 & 2 & $-1$ & $-1$  \\
        $U_{L,R}$ & 3 & 1 & 1 & $4/3$ & $1/3$  \\
        $D_{L,R}$ & 3 & 1 & 1 & $-2/3$ & $1/3$  \\
        $E_{L,R}$ & 1 & 1 & 1 & $-2$ & $-1$  \\
        \hline
        $H_{L}$ & 1 & 2 & 1 & 1 & 0  \\
        $H_{R}$ & 1 & 1 & 2 & 1 & 0  \\
        $\Phi_1$ & 1 & 1 & 1 & 0 & $q$  \\
        $\Phi_2$ & 1 & 1 & 1 & 0 & $2q$  \\
        \hline
    \end{tabular}
    \caption{Charge assignment for fermions and scalars under $SU(3)_C\times SU(2)_L\times SU(2)_R\times U(1)_X \times U(1)_{B-L}$.
    }
    \label{tab:charge}
\end{table}
With these particle content, the most general renormalizable Yukawa Lagrangian of the model is given by
\begin{align}
\label{eq:LaYuk}
    \mathcal{L}_Y =& ~~y_u\ (\bar{Q}_L \tilde{H}_L + \bar{Q}_R \tilde{H}_R) U + y_d\ (\bar{Q}_L H_L + \bar{Q}_R H_R) D \nonumber\\
    &+ y_\ell\ (\bar{\Psi}_L H_L + \bar{\Psi}_R H_R) E + h.c. 
\end{align}
where $\tilde{H}_{L,R} = i \tau_2 H^*_{L,R}$.  Here we imposed parity symmetry under which $Q_{L} \leftrightarrow Q_{R}$, $\psi_{L} \leftrightarrow \psi_{R}$, $F_L \leftrightarrow F_R$, and $H_L \leftrightarrow H_R$. In addition, the theory also includes bare mass terms for vector-like fermions (VLF) $M_F \bar{F}F$. 
After the symmetry breaking, the Lagrangian terms of Eq.~\eqref{eq:LaYuk} together with VLF masses lead to $6\times 6$ mass matrices for up-type quarks $(u,U)$, down-type quarks $(d,D)$ and charged leptons $(e,E)$:
\begin{align}
    {\cal M}_f = 
    \begin{pmatrix}
    0 & y_f \,\kappa_L \\
    y_f^\dagger\, \kappa_R & M_{F}
    \end{pmatrix} \, ,
    \label{eq:massfermion}
\end{align}
where $f$ stands for $\{u,d,\ell\}$. The above mass matrices can be block-diagonalized to obtain the light fermion mass matrices: 
\begin{equation}
    M^{\rm light}_f \simeq - y_f M_F^{-1} y_f^\dagger \kappa_L \kappa_R \, ,
    \label{eq:fermionmass}
\end{equation}
where we have assumed  $M_F \gg y_f \kappa_R, y_f \kappa_L$. In this generalized seesaw mechanism, a milder hierarchy in the Yukawa couplings ($y_i = 10^{-3} - 1$) is sufficient to explain the mass hierarchy of charged fermions, contrary to the SM setup which spans over the range $(10^{-6}-1)$.   

\begin{figure}
    \centering
    \includegraphics[width=7cm,height=3.5cm]{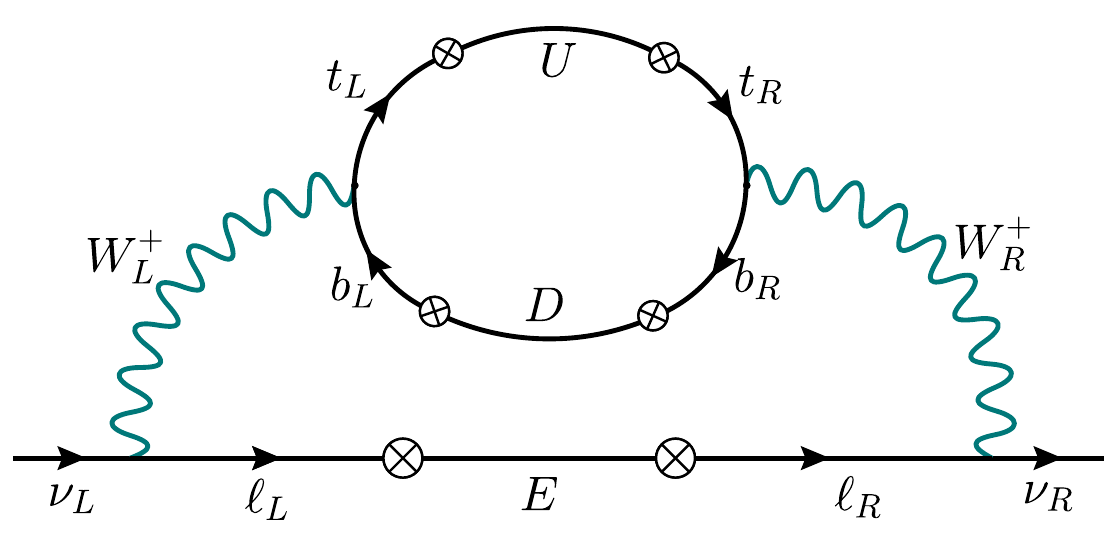}
    \caption{Two-loop radiative Dirac neutrino masses.}
    \label{fig:numass}
\end{figure}

Note that the neutrino masses are not generated at the tree-level via this generalized seesaw scheme due to the absence of vector-like neutral fermions. Instead, their masses arise at the quantum level via two-loop radiative diagrams shown in Fig.~\ref{fig:numass}. The corresponding contributions to neutrino mass are given by \cite{Babu:1988yq,Babu:2022ikf} 
\begin{align}
    M_D \simeq & -\frac{g^2 m_t m_b y_\ell M_E}{(16\pi^2)^2 m_{W_L}m_{W_R}} \  G_2\left[ \frac{M_D^2}{M_E^2}, \frac{M_U^2}{M_E^2} \right] y_\ell^\dagger \, ,
\end{align}
where $g_L=g_R=g$ and function $G_2[r_1,r_2]$ is given explicitly in Ref.~\cite{Babu:2022ikf}. 
It is important to note that these two-loop contributions only induce Dirac mass terms for neutrinos with no contributions to the Majorana masses even at the loop level \cite{Babu:2022ikf}.  However, they can be generated via higher dimensional Planck-suppressed operators, a critical step to realize pseudo-Dirac neutrinos in our framework, which we discuss next.

\begin{figure}
    \centering
\includegraphics[width=7.8cm,height=7.2cm]{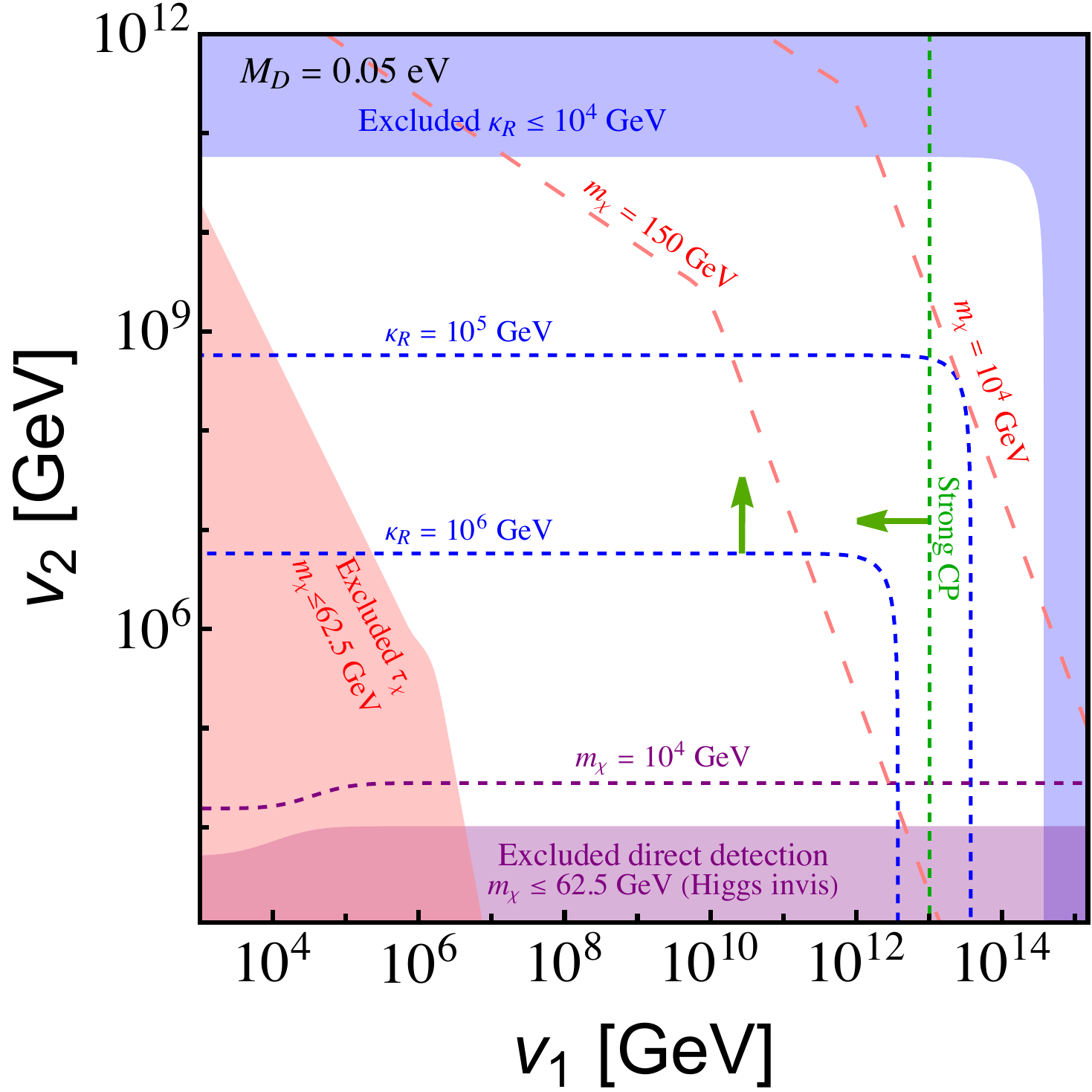}
    \caption{Limit on the $U(1)_{B-L}$ breaking VEVs $v_1$ and $v_2$. Blue contour represents the upper limit for a fixed $SU(2)_R$ breaking VEV $\kappa_R$ obtained from $\delta m^2 \lesssim 10^{-11}$ eV$^2$. Purple contour corresponds to the lower limit on $v_2$ for a fixed $m_{\chi}$ derived from the direct detection. Red contour represents the lower limit on $(v_1, v_2)$ for a fixed $m_\chi$ obtained from pNGB DM lifetime. The mass $m_\chi \lesssim 62.5$ GeV is excluded from Higgs invisible decay. The green arrow indicates the parameter space consistent with the Strong CP solution. }
    \label{fig:v1v2plot}
\end{figure}

\medskip
\textbf{\emph{Realizing pseudo-Dirac neutrinos}.--}
The contributions to the Majorana mass terms arise via the following operators in our model:
\begin{align}
        {\cal O}_{5+k;(R,L)} \equiv &\   \frac{1}{M_{\rm Pl}^{k+1}}\ \Psi_{R,L} \Psi_{R,L} H_{R,L} H_{R,L} \Phi_i^k
        \label{eq:Planckop}
\end{align}
with the  charge assignment $q=2/k$ for $\Phi_1$ and $q=1/k$ for $\Phi_2$.  Here we take $q=1$ such that $d=6$ ($d=7$) operator is induced for $\Phi_2$ ($\Phi_1$). Taking into account these contributions, the neutrino mass matrix in the $(\nu, \nu^c)$ basis can be written as
\begin{equation}
\mathcal{M}=\left(\begin{array}{ll}M_{L} & M_{D} \\ M_{D}^T & M_{R}\end{array}\right),
\end{equation}
where $M_{L,R}$ correspond to the Majorana masses obtained from the Planck-induced operators given in Eq.~\eqref{eq:Planckop}. 
The corresponding eigenvalues are given by
\begin{equation}
   m_{\pm}= \frac{1}{2} \left(M_L+M_R \pm \sqrt{4 M_D^2+\left(M_L-M_R\right){}^2}\right).
   \label{eq:neutEigenValues}
\end{equation}
The mass term $M_L$ can be safely ignored here as it is of order $M_R \kappa_L/\kappa_R$. 
To realize pseudo-Dirac neutrinos, $M_D \gg M_R, M_L$ is required which is naturally satisfied due to Planck-suppressed operators inducing $M_{L,R}$. The mass-squared difference between the active-sterile neutrino states is then given as $\delta m^{2}=m_{+}^{2}-m_{-}^{2}  \simeq 2 M_{D} M_{R}$. 
The constraint on this mass-splitting from solar neutrino data, $\delta m^2 \lesssim 10^{-11}\ \text{eV}^2$ \cite{deGouvea:2009fp,Chen:2022zts,Ansarifard:2022kvy}, puts a stringent limit on the $U(1)_{B-L}$ breaking scale $v_{i}$ for a fixed $SU(2)_R$ breaking scale $\kappa_R$. Assuming coefficients of $d=6, 7$ operators are of order one and equal, we find
\begin{align}
\label{eq:delta_m2_appr}
\delta m^2 \simeq \dfrac{2\kappa_R^2 M_D}{M_{Pl}^2}\left(v_2 + \dfrac{v_1^2}{M_{Pl}}\right) \, .
\end{align}
We then recast it to obtain upper limits on $v_{1,2}$ for different choices of $\kappa_R$, shown by the blue contours in Fig.~\ref{fig:v1v2plot}. Note that $\kappa_R \lesssim 10$ TeV is disfavored from resonance searches of new neutral gauge bosons \cite{ATLAS:2019erb}, assuming relevant gauge coupling strength of the same order as of weak interaction strength.

The $B-L$ gauge symmetry is crucial in realizing a consistent scenario of pseudo-Dirac neutrinos. In models without gauged $U(1)_{B-L}$ symmetry, the Planck-suppressed operators arise at $d=5$,  $\Psi_{R} \Psi_{R} H_{R} H_{R}/M_{\rm Pl}$, as the quantum gravitational corrections are not expected to respect global symmetries \cite{Banks:2010zn}. This results in a Majorana mass $M_R \gtrsim {\cal O} (10^{-2})$ eV  (taking $\kappa_R  \gtrsim 10$ TeV with order one coefficient), yielding $\delta m^2 \sim 10^{-3}$ eV$^2$- significantly exceeding the allowed values from cosmology \cite{Barbieri:1989ti,Enqvist:1990ek} and neutrino data \cite{deGouvea:2009fp}. However, gauging the $B-L$ symmetry forbids such $d=5$ operators, as the quantum gravitational interactions are expected to preserve local symmetries.

\medskip
\textbf{\emph{A viable candidate for pNGB DM}.--}
We turn our attention to realizing pNGB DM candidate within the model by first analyzing the scalar potential
\begin{align}
\label{eq:pot}
    V =& -m_{H_L}^2 H_L^\dagger H_L -m_{H_R}^2 H_R^\dagger H_R  - m_{\Phi_i}^2 \Phi_i^\dagger \Phi_i    \notag \\
   & + \lambda_H H_L^\dagger H_L H_R^\dagger H_R + \lambda'_H \{ (H_L^\dagger H_L)^2 + (H_R^\dagger H_R)^2 \}   \notag \\
   &+ \lambda_{\Phi_i} (\Phi_i^\dagger \Phi_i)^2 +\lambda_{H\Phi_i} \Phi_i^\dagger \Phi_i \{ H_L^\dagger H_L + H_R^\dagger H_R \}  \notag \\
   & + \lambda'_{\Phi} (\Phi_1^\dagger \Phi_1)(\Phi_2^\dagger \Phi_2) -\sqrt{2} \mu\  (\Phi_2^\dagger \Phi_1^2+ \Phi_2 \Phi_1^{\dagger2})  \, .
\end{align}
Here $i=1,2$ and all parameters are real. Parity symmetry is softly broken by the condition $m_{H_L}^2\neq m_{H_R}^2$ to achieve $\kappa_R \gg \kappa_L$, guaranteeing the required mass splittings between left- and right-handed gauge bosons \cite{Mohapatra:1974gc}. This accounts for parity non-conservation and dominant $V-A$ character of known weak interactions.

The potential remain invariant under the $\Phi_{i} \rightarrow \Phi_{i}^{\dagger}$ implying an accidental $Z_2$ symmetry. A real (imaginary) component of the field $\Phi_{1,2}$ is $Z_2$-even (-odd) under such transformation. When $\Phi_{1,2}$ develop VEVs and break the $B-L$ symmetry, a linear combination of the fields ($\Im[\Phi_1]$, $\Im[\Phi_2]$) orthogonal to the Goldstone mode would constitute a viable candidate for pNGB DM. The mass of this physical state is proportional to the cubic coupling $\mu$ (cf. Eq.~\eqref{eq:massDM}). Note that the covariant derivatives of $\Phi_{1,2}$ break this $Z_2$ symmetry, allowing pNGB DM to decay via gauge interactions. 

Let us investigate the scalar sector using the following parameterization for the scalar multiplets:
\begin{align}
\label{eq:parametrize}
   H_{L,R} &= \begin{pmatrix}
   G_{L,R}^{+}\\
   (h_{L,R}+\kappa_{L,R}+ i\ G^0_{L,R})/\sqrt{2}
   \end{pmatrix}, \\ 
    \Phi_{1,2}&= \left(\phi_{1,2}+v_{1,2}+i\ \chi_{1,2}\right)/\sqrt{2}. 
\end{align}
Here $G_{L,R}^+$ ($G_{L,R}^0$) are the Goldstone modes absorbed by $W_{L,R}^+$ ($Z_{L,R}$) gauge bosons. After spontaneous symmetry breaking and solving stationary conditions, the symmetric mass matrix for the CP-even scalar in the $(h_L, h_R, \phi_1, \phi_2)$ basis reads as ${\cal M}_H^2=$
\begin{align}   
& \begin{pmatrix}
  2\lambda'_H\kappa_L^2 &  \lambda_{H}\kappa_L\kappa_R & \lambda_{H\Phi_1}v_1\kappa_L & \lambda_{H\Phi_2}v_2\kappa_L\\
  \bullet &  2\lambda'_H\kappa_R^2 & \lambda_{H\Phi_1}v_1\kappa_R & \lambda_{H\Phi_2}v_2\kappa_R\\
  \bullet & \bullet & 2 \lambda_{\Phi_1}v_1^2 & v_1\left(\lambda'_{\Phi}v_2-2\mu\right)\\
  \bullet & \bullet & \bullet & 2\lambda_{\Phi_2}v_2^2+\dfrac{\mu v_1^2}{v_2} 
  \label{eq:realScalar}
\end{pmatrix} ,
\end{align}
which can be diagonalized by
\begin{equation}
    U {\cal M}_H^2 U^T = {\cal M}_H^{2, {\rm diag}} \, .
    \label{eq:unitaryReal}
\end{equation} 
Here $U$ represents the orthogonal matrix that transforms the original basis $(h_L, h_R, \phi_1, \phi_2)$ into the mass basis $(h_1,h_2, h_3, h_4)$. We identify $h_1$ to be SM-like Higgs. As for the CP-odd scalars, the mass eigenstates  $(G', \chi)$ with masses 
\begin{equation}
    m_{G'}^2 = 0, \hspace{5mm} m_{\chi}^2 = \mu (v_1^2+4v_2^2)/v_2 \, ,
    \label{eq:massDM}
\end{equation}
are related to $(\chi_1, \chi_2)$ via the following transformations
\begin{align}
    G' = \cos\theta \chi_1 + \sin\theta \chi_2 \, , \hspace{3mm}
    \chi = -\sin\theta \chi_1 + \cos\theta \chi_2 \, ,
\end{align}
where the mixing angle $\theta$ given by 
\begin{equation}
    \sin 2\theta = \frac{4 v_1 v_2}{v_1^2 + 4 v_2^2} \, .
    \label{eq:theta}
\end{equation}
Here $G'$ is the Goldstone mode absorbed by the gauge boson associated with the $B-L$ gauge symmetry and the scalar state $\chi$ is identified as the pNGB DM.

\medskip
\textbf{\emph{Suppression of direct detection amplitude and relic abundance of DM}.--}
In this section, we confirm that the spin-independent cross-section of the pNGB DM with nucleon is indeed suppressed in the zero momentum transfer limit. The pNGB DM $\chi$ interacts with nucleons via the $t$-channel exchange of CP-even scalars. The elastic scattering amplitude in the interaction basis reads as
\begin{align}
    {\cal A} \propto C_{\chi \chi H}\ (M_H^2)^{-1}\ C_{H f f}^T \, ,  
\end{align}
where $C_{H f f} = M_f(1/\kappa_L, 1/\kappa_R, 0, 0)$ is the effective coupling of CP-even scalars $H=(h_L, h_R, \phi_1, \phi_2)$ with SM fermions. $M_H^2$ and $M_f$ are given in Eqs.~\eqref{eq:realScalar} and \eqref{eq:fermionmass}. $C_{\chi \chi H} = (C_L, C_R, C_1, C_2)$ is the effective couplings of $\chi$-$\chi-H$ and read as
\begin{align}
    C_{L,R} &= \kappa_{L,R} \left(\lambda_{H\Phi_2}v_1^2+4\lambda_{H\Phi_1}v_2^2\right)/(2\left(v_1^2+4v_2^2\right)) , \label{eq:CL} \notag \\
    C_1 &= v_1\left(\lambda_{\Phi}'v_1^2+8v_2(\mu+\lambda_{\Phi_1}v_2)\right)/(2\left(v_1^2+4v_2^2\right)), \notag \\
    C_2 &= v_2\left(\lambda_{\Phi_2}v_1^2+2v_2(2\mu+\lambda_{\Phi}'v_2)\right)/\left(v_1^2+4v_2^2\right) .
\end{align}
In the limit $\kappa_L \ll \kappa_R$ and $m_{\chi} \ll v_1, v_2$, the amplitude takes the following form 
\begin{align}
\label{eq:DDAmpl-2}
\mathcal{A}\propto \frac{M_f m_{\chi}^2\left(\Lambda _1 v_1^2+\Lambda _2 v_2^2\right)}{2 \Lambda _{3} \kappa_L^2 v_2^2 \left(v_1^2+4 v_2^2\right)} \propto \frac{M_f m_{\chi}^2}{\kappa_L^2 v_2^2 } \, .
\end{align}
Here $\Lambda_i$ are some combination of the quartic couplings given in Eq.~\eqref{eq:pot} and can be of order one. 
This is adequately suppressed and gives a constraint on $U(1)_{B-L}$ breaking scale $v_2$ for a fixed DM mass. 
We estimate the spin-independent DM-nucleon cross-section~\cite{LZ:2022lsv}:
\begin{align}
    \sigma_{\rm SI} \simeq \dfrac{A^2}{\pi} \frac{m_\chi^2}{ \kappa_L^4 v_2^4} m_N^4 f_N^2,
    \label{eq:DDCross}
\end{align}
where $A=Z+N$ refers to the nucleon number, $m_N$ is the nucleon mass and $f_N\approx 0.3$ describe the effective Higgs-nucleon interaction~\cite{Hoferichter:2017olk}. We used full expression for the amplitude to obtain the lower limit on $v_2$ from Lux-Zeplin \cite{LZ:2022lsv}, depicted in Fig.~\ref{fig:v1v2plot} by purple contour for a fixed DM mass, $m_{\chi}= 62.5$ GeV and $10^4$ GeV.
\begin{figure}
    \centering
\includegraphics[width=8cm, height=5.5cm]{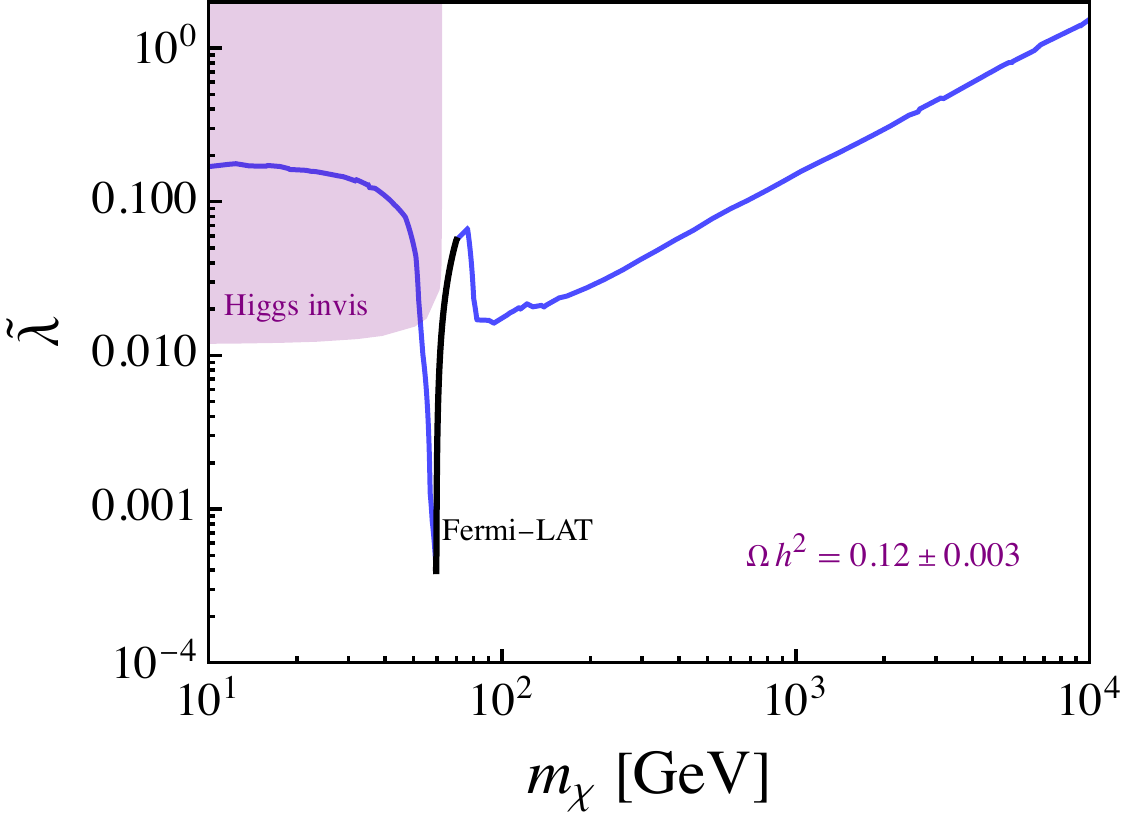}
    \caption{Allowed regions in the $(m_\chi, \tilde{\lambda})$ plane that satisfies relic abundance requirement. The purple region is excluded by the Higgs invisible decay \cite{CMS:2018yfx}. The black segment along the abundance curve is excluded by Fermi-LAT+MAGIC bound~\cite{MAGIC:2016xys}.
    }
    \label{fig:darkmatter}
\end{figure}

Since one can evade direct detection constraints in the model, a vast region of parameter space opens up for the DM relic abundance.  
In the natural limit when the quartic couplings are of ${\cal O} (1)$, a hierarchy in the Higgs sector is realized such that the only relevant interaction of DM  annihilation into SM particles is through SM-like Higgs boson $h_1$.
The relevant interaction is given by
\begin{equation}
        V \supset \tilde{\lambda}/2\  \chi^2  h_1^2 + \tilde{\lambda} \kappa_L\ \chi^2 h_1 \, ,
    \end{equation} 
where $\tilde{\lambda} \simeq  C_L/\kappa_L$ with $C_L$ defined in Eq.~\eqref{eq:CL}. We implement the model with these couplings in {\tt CalcHEP} \cite{Belyaev:2012qa} by using {\tt LanHEP} \cite{Semenov:2014rea} and numerically evaluate the relic abundance of the pNGB DM using the software {\tt MicrOmegas} \cite{Belanger:2014vza}. The allowed parameter space is shown in Fig.~\ref{fig:darkmatter} where the blue curve satisfies DM relic density $\Omega h^2 = 0.120 \pm 0.003$  \cite{Planck:2018vyg}. For the DM mass below $m_{h_1}/2$, the SM-like Higgs $h_1$ can decay to a pair of pNGB DMs. Such invisible decay is constrained to be BR$(h_1 \to \chi \chi) < 0.16$ \cite{CMS:2018yfx}, depicted in Fig.~\ref{fig:darkmatter} by purple-shaded region. 
Additionally, indirect detection experiments such as Fermi-LAT \cite{Fermi-LAT:2015att} and MAGIC \cite{MAGIC:2016xys} impose limits on the model parameter space shown by the black shaded segment in Fig.~\ref{fig:darkmatter}.

\medskip
\textbf{\emph{Lifetime of dark matter and its connection with  \pmb{$\delta m^2$}}.--}
As aforementioned, the covariant derivatives of $\Phi_{1,2}$ break the $Z_2$ symmetry, resulting in the decay of the DM. Therefore, one has to ensure that it is sufficiently long-lived $\tau_\chi \gtrsim 10^{27}$ s \cite{Baring:2015sza}. The DM candidate in the model can have two-, three-, and four-body decay modes. 

The two-body decay process $\chi \to h_i Z$ can be particularly important when $m_{\chi}>m_Z+m_{h_{i}}$. The total decay width for this decay channel is given by $\Gamma_{\chi \to Z h_i} \simeq$ 
\begin{align}
\label{eq:DMdecay_2}
  \dfrac{\xi^2 g_{B}^2 m_\chi}{16\pi }  \left(2 c_\theta  U_{4i} - s_\theta U_{3i}\right)^2 {\cal F}_1 \left[ \frac{m_{Z}^2}{m_\chi^2}, \frac{m_{h_i}^2}{m_\chi^2}\right] \, ,
\end{align}
where ${\cal F}_1 [a, b] \equiv  1/a \left\{ a^2 + (b-1)^2 -2 a (b+1) \right\}^{3/2}$.
Here $(s_{\theta}\equiv \sin\theta,c_{\theta}\equiv \cos\theta)$ is defined in Eq.~\eqref{eq:theta} and $U$ is the orthogonal matrix of Eq.~\eqref{eq:unitaryReal}. $\xi$ denotes the mixing angle between $B^{\mu}$ and $Z_L^{\mu}$, which is related to the gauge kinetic mixing $\epsilon$; $ \xi = \tan\epsilon \, g_X m_{Z}^2 \kappa_L^2/m_{Z''}^4$.
Notice that this decay mode becomes insignificant for $\epsilon \ll 1$ and larger masses of $Z''$. 

In the case when the two-body decay channels are kinematically forbidden or suppressed, the three-body decay process $\chi \to Z f \bar{f}$ and $\chi \to h_i f \bar{f}$, though phase-space suppressed, may become important. The contribution to $\Gamma_{\chi}$ from $\chi \to Z f \bar{f}$ is negligible as it is suppressed by the heavy mass of $h_3,h_4$ and small scalar mixing. Additionally, when $\epsilon \to 0$ this decay mode vanishes. Conversely, the latter process contributes significantly providing stringent bound on the model parameter space.
In the limit $M_{f} \ll m_{h_i}, m_{\chi} \ll m_{Z''}$ and $\epsilon = 0$, the decay-width to this process is given by \cite{Abe:2020iph}: 
\begin{align}
 \Gamma_{\chi \to h_i f \bar{f}} \simeq  \frac{g_{B}^4 m_{\chi}^5 \left(2 c_\theta  U_{4i} - s_\theta U_{3i}\right)^2}{  768 \pi^3 m_{Z''}^4}  {\cal F}_2\left[ \frac{m_{h_i}^2}{m_\chi^2} \right] ,
\end{align}
where ${\cal F}_2 [a] =  1 - 8a + 8 a^3 -a^4 -12 a^2 \log a$ .
The four-body decay channel $\chi \to h_i^* B^* \to f \bar{f}f \bar{f}$ is particularly important when $m_\chi < m_h$.
We estimate its contribution to $\Gamma_{\chi}$: 
\begin{align}
   \Gamma_{\chi\to 4f} \approx &\ \frac{2 g_B^4 m_\chi^5}{(10 \pi)^5 m_{Z''}^4}  \left[ s_\theta^2 \frac{M_b^2 m_\chi^4 }{4 v_1^2 m_{h_3}^4} + c_\theta^2 \frac{M_b^2 m_\chi^4 }{v_2^2 m_{h_4}^4}  \right] \, .
\end{align}
Here $M_b$ is the mass of the bottom quark.

Taking into account all the decay processes and fixing DM lifetime $\tau_\chi \gtrsim 10^{27}$ s \cite{Baring:2015sza}, we obtain the lower limit on $(v_1, v_2)$ which is illustrated by the red contours in Fig.~\ref{fig:v1v2plot}. To adequately suppress $\Gamma_{\chi}$ one requires larger values of $v_{1,2}$. However, increasing values of $v_{1,2}$ also significantly impact $\delta m^2$, see Eq.~\eqref{eq:delta_m2_appr}. Thus through this reciprocal dependency on the \(U(1)_{B-L}\) breaking scale, the lifetime of DM and $\delta m^2$ are strongly correlated within our setup.
\begin{figure}
    \centering    \includegraphics[width=8cm, height=7cm]{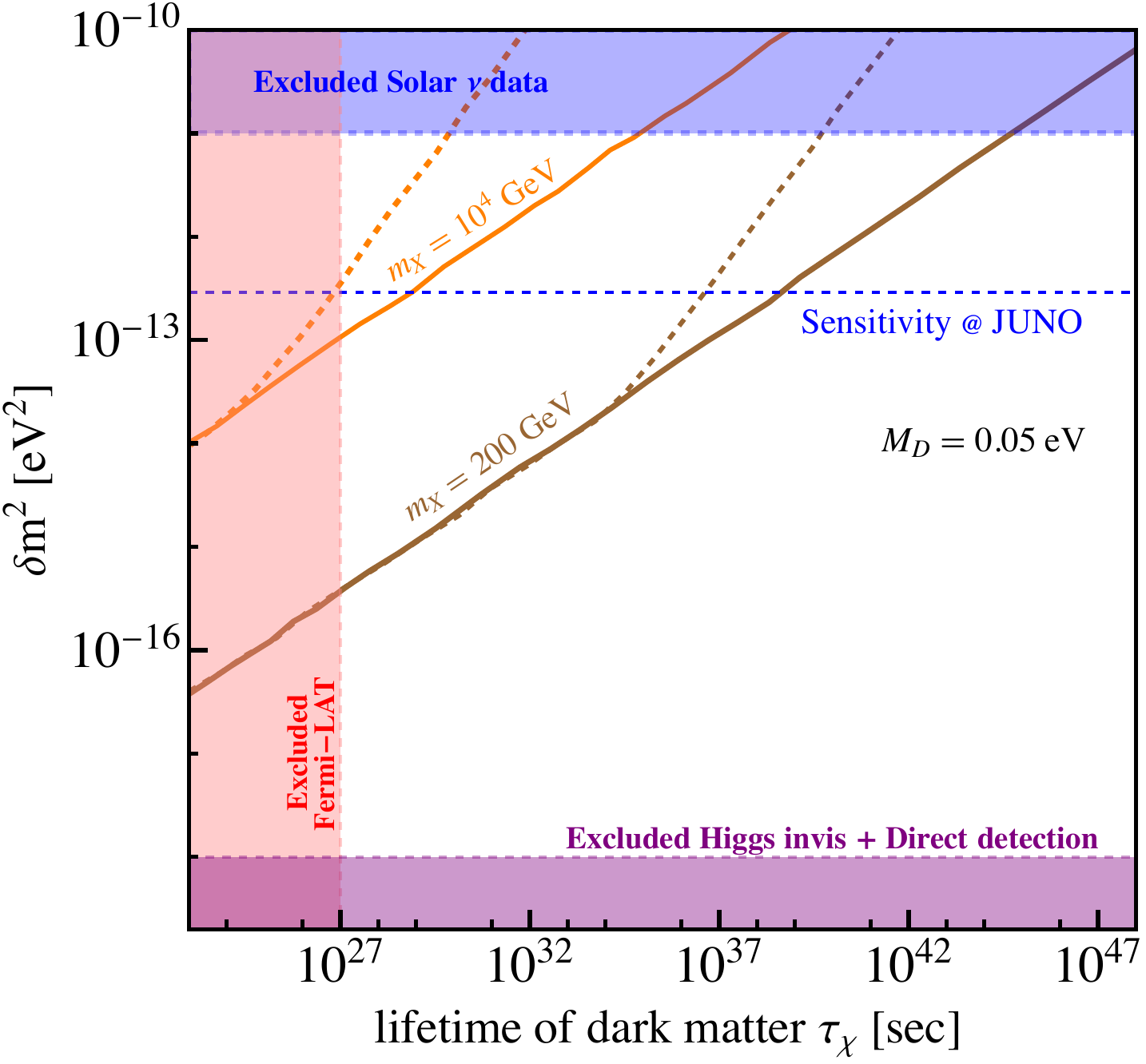}
    \caption{Allowed parameter space in the lifetime of pNGB DM  $\tau_\chi$ and the neutrino mass-squared splitting $\delta m^2$. Brown and orange solid (dashed) lines represent the lower bound for $200$ GeV and $10^4$ GeV DM masses without (with) the Strong CP solution. Purple shaded region is the exclusion limit from Higgs invisible decay, $h_1 \to \chi \chi$ together with limits from Direct detection on $v_2$. Blue dashed line represents the future sensitivity reach of the JUNO experiment \cite{Franklin:2023diy}.}
    \label{fig:moneyplot}
\end{figure}

We illustrate this direct correlation for different DM masses in Fig.~\ref{fig:moneyplot}. The plot is generated by fixing the DM mass and varying $v_1$, $v_2$ and $\kappa_R$ within the allowed range set by DM lifetime constraint (lower-bound) and the limit  $\delta m^2 \lesssim 10^{-11}\ \text{eV}^2$~\cite{deGouvea:2009fp,Chen:2022zts,Ansarifard:2022kvy} (upper-bound). Additionally, direct detection experiments also put a lower bound on $v_2$. The region above each contour line denotes the allowed parameter space, while the area below indicates the absence of a viable solution for any choice of $v_{1,2}$. Color-shaded regions depict excluded parameter spaces: red for DM lifetime and blue for $\delta m^2$ limits.  
It is clear that higher DM masses predict larger $\delta m^2$, as they require larger $v_{1,2}$ to satisfy the lifetime of DM constraint. When $m_\chi < m_h$,  the relevant mode is the 4-body decay that easily satisfies the lifetime constraint. Direct detection set a lower limit on $v_2$  (cf. Fig~\ref{fig:v1v2plot}) which in turn sets a lower limit on $\delta m^2$, depicted by purple band in Fig.~\ref{fig:moneyplot}. 

\medskip
\textbf{\emph{Parity solution to Strong CP}.--}
Our framework allows for a parity solution to the Strong CP problem. The physical invariant parameter that contributes to the neutron EDM is given by $\bar{\theta} = \theta_{\rm QCD} + {\rm Arg}\{ {\rm det} ({\cal M}_q) \}$, 
where ${\cal M}_q$ is the quark mass matrix of Eq.~\eqref{eq:fermionmass}. $\theta_{\rm QCD}$ vanishes in parity symmetric theory and ${\rm Arg}\{ {\rm det} ({\cal M}_q) \}$ is real. Thus $\bar{\theta} = 0$ at tree-level. Nonzero $\bar{\theta}$ can arise from radiative corrections, potentially within the experimentally allowed range $\bar{\theta} \leq 1.19 \times 10^{-10}$ \cite{Dragos:2019oxn}. It has been shown that $\bar{\theta}$ vanishes at one-loop order and very small $\bar{\theta}$ would be induced through two-loop radiative corrections \cite{Babu:1989rb}. 
Notice that the Planck-induced corrections would allow $d=5, 6$ operators of the type 
\begin{align}
    \mathcal{O}_5&\equiv (\bar{Q}_L Q_R) H_R^\dagger H_L /M_{\rm Pl},\\
    \mathcal{O}_6 &\equiv \bar{Q}_{L,R}  U_{R,L}\tilde{H}_{L,R}\Phi_i^{\dagger}\Phi_i/M_{\rm Pl}^2 \, ,
\end{align}
which induce non-hermitian entries in the quark mass matrices, thus affecting the Strong CP solution. Demanding $\bar{\theta} \leq 10^{-10}$ puts an upper limit on both the left-right and $U(1)_{B-L}$ breaking scale, $\kappa_R \leq 10^6$ GeV and $v_{1,2} \leq 10^{13}$ GeV. This further constrains the model parameter space as shown in Fig.~\ref{fig:v1v2plot} and Fig.~\ref{fig:moneyplot}.  

\medskip
\textbf{\emph{Conclusions}.--}
We proposed a realization of pNGB DM within LRSM, wherein the neutrinos are naturally pseudo-Dirac. The DM candidate arises from $\Phi_{1,2}$, both charged under a $U(1)_{B-L}$ gauge symmetry. Unlike conventional Higgs-portal scalar DM, the proposed model naturally evades stringent constraints from direct detection due to its Nambu-Goldstone boson nature. This in turn opens up wide region of parameter space $\mathcal{O}(10)\, \mathrm{GeV}\leq m_{\chi}\leq \mathcal{O}(10)\, \mathrm{TeV}$ to satisfy relic abundance. Since the DM in the model can decay, consistency from lifetime constraints necessitates the larger VEVs of $\Phi_{1,2}$. Interestingly, the mass-squared difference $\delta m^2$ between the active and sterile neutrinos also depends on these VEVs via Planck-suppressed operators. We have thus shown that $\delta m^2$ and DM lifetime are directly correlated within the model through these reciprocal dependencies on the $U(1)_{B-L}$ breaking scales while maintaining the parity solution to the Strong CP problem. 

\begin{acknowledgments}
{\textbf {\textit {Acknowledgments.--}}} The work of AT was supported in part by the National Science Foundation under Grant PHY-2210428. The work of SB is supported in part by the U.S. Department of Energy under grant number DE-SC0016013. 
\end{acknowledgments}

\onecolumngrid
\appendix

\section{Neutral gauge bosons sector} \label{App:gaugesector}
\noindent
The neutral gauge boson associated with the symmetry $SU(2)_L$, $SU(2)_R$, $U(1)_X$, and $U(1)_{B-L}$ is represented by $W_{\mu L}^0$, $W_{\mu R}^0$, $X_\mu $, and $B_\mu $, respectively.  These gauge fields will mix each other and produce physical gauge bosons $A_\mu,Z_{\mu L},Z_{\mu}',$ and $Z_\mu''$ states, where $A_\mu$ and $Z_{\mu L}$ can be identified as the SM  gauge fields $A_\mu$ and $Z_\mu$. The corresponding mass-squared matrix in the basis    $(W_{\mu L}^0, W_{\mu R}^0, X_\mu, B_\mu)$ reads as
\begin{equation}
M_0^2=
\begin{pmatrix}
        \frac{1}{4}g_L^2\kappa_L^2& 0 & -\frac{1}{4}g_Xg_L\kappa_L^2 & 0\\
        0 & \frac{1}{4}g_R^2\kappa_R^2 & -\frac{1}{4}g_X g_R\kappa_R^2 & 0 \\
        \bullet & \bullet & \frac{1}{4}g_X^2(\kappa_L^2+\kappa_R^2) & 0\\
        \bullet & \bullet & \bullet & g_B^2(v_1^2+4v_2^2)
\end{pmatrix}.
\end{equation}
The above mass matrix gets modified in the presence of the gauge kinetic mixing between the fields $X_\mu$ and $B_\mu$, $ -\frac{\sin \epsilon}{2} B_{\mu\nu} X^{\mu\nu} $, where
$B_{\mu\nu}$ ($X_{\mu\nu}$) is the field strength tensor associated with gauge boson $B_\mu$ ($X_\mu$). In the basis where gauge kinetic terms are diagonalized, indicated by $ (W_{\mu L}^0, W_{\mu R}^0, \hat{X_\mu},\hat{B_\mu})$, the mass-squared matrix takes the following form
\begin{align}
\Tilde{M}_0^2 =\left(
\begin{array}{cccc}
 \frac{1}{4} g_L^2 \kappa _L^2 & 0 & -\frac{1}{4} g_L g_X \kappa _L^2 & \frac{1}{4} g_L g_X \kappa _L^2 \tan \epsilon \\
 0 & \frac{1}{4} g_R^2 \kappa _R^2 & -\frac{1}{4} g_R g_X \kappa _R^2 & \frac{1}{4} g_R g_X\kappa _R^2 \tan\epsilon \\
 \bullet & \bullet & \frac{1}{4} g_X^2 \left(\kappa _L^2+\kappa _R^2\right) & -\frac{1}{4} g_X^2\left(\kappa _L^2+\kappa _R^2\right)\tan \epsilon \\
 \bullet & \bullet & \bullet & g_B^2\left(v_1^2+4 v_2^2\right)  \sec ^2\epsilon+\frac{1}{4} g_X^2 \left(\kappa _L^2+\kappa _R^2\right) \tan ^2\epsilon\\
\end{array}
\right).
\label{eqn:rm}
\end{align}
The states $(W_{\mu L}^0, W_{\mu R}^0, \hat{X_\mu},\hat{B_\mu})$ will mix to produce $(A_\mu, Z_{\mu L}, Z_{\mu R},  B_{\mu}^{\prime})$, in analogy with the SM. The photon field $A_\mu$ remains massless, while the rest of the fields mix. It is convenient to choose the following basis: 
\begin{align}
A_{\mu}&=\frac{g_X g_R W_{\mu L}^0 + g_X g_L W_{\mu R}^0 + g_L g_R \hat{X_\mu}}{\sqrt{g_X^2 (g_R^2 + g_L^2) + g_L^2 g_R^2}} \, , \hspace{10 mm}  Z_{\mu L}=\dfrac{g_L(g_R^2+g_X^2) W_{\mu L}^0  - g_R g_X^2 W_{\mu R}^0 - g_R^2g_X \hat{X_\mu} }{\sqrt{(g_R^2+g_X^2)(g_R^2g_X^2+g_L^2(g_R^2+g_X^2))}} \notag \\
Z_{\mu R}&=\dfrac{-g_R W_{\mu R}^0 + g_X \hat{X_\mu} }{\sqrt{g_R^2+g_X^2}}, \hspace{35 mm} B_{\mu}^{\prime}=\hat{B_\mu}.
 \label{eqn:gauge_eigen}
\end{align}
In this basis the photon field decouples from other gauge fields, leading to a $3\times 3$ mass-squared matrix $B^2$ reads as
\begin{equation}
B^2=\left(
\begin{array}{ccc}
 \dfrac{1}{4} \left(g_L^2+\dfrac{g_R^2 g_X^2}{g_R^2+g_X^2}\right)\kappa _L^2 & -\dfrac{g_X^2\sqrt{g_L^2 \left(g_R^2+g_X^2\right)+g_R^2 g_X^2}}{4 \left(g_R^2+g_X^2\right)}\kappa _L^2 & g_X\sqrt{\dfrac{g_L^2 \left(g_R^2+g_X^2\right)+g_R^2 g_X^2}{16(g_R^2+g_X^2)}}\kappa _L^2 \tan \epsilon \\
\bullet & \dfrac{g_X^4 \kappa _L^2+ \left(g_R^2+g_X^2\right)^2\kappa _R^2}{4 \left(g_R^2+g_X^2\right)} & -\dfrac{g_X\left(g_X^2 \left(\kappa _L^2+\kappa _R^2\right)+g_R^2 \kappa _R^2\right)}{4 \sqrt{g_R^2+g_X^2}}\tan\epsilon  \\
\bullet& \bullet &  g_B^2 \left(v_1^2+4 v_2^2\right)\sec ^2\epsilon+\dfrac{1}{4} g_X^2 \left(\kappa _L^2+\kappa _R^2\right)\tan ^2\epsilon  \\
\end{array}
\right).
\end{equation}
As one can see from the above equation, in the $\kappa_L\rightarrow 0$ limit, the eigenvalue associated with $Z_{\mu L}$ is zero, which resembles the SM $Z_\mu$-boson. With the hierarchy $\kappa_R\ll v_1, v_2$, the gauge fields $Z_{\mu L},~Z_{\mu R}$, and $B_\mu'$ from equation~\eqref{eqn:gauge_eigen} can be approximated as $Z_\mu,Z_\mu'$ and $Z_\mu''$ with the eigenvalues,
\begin{align}
    M_{Z}^2&= \dfrac{1}{4} \kappa _L^2 \left(g_L^2+g_Y^2\right)+\mathcal{O}\left(\dfrac{\kappa_L^2}{\kappa_R^2}\right),
    \label{eq:gaugemass1}\\
    M_{Z'}^2&=\dfrac{1}{4}\dfrac{g_Y^4 \kappa _L^2+g_R^4 \kappa _R^2}{g_R^2-g_Y^2}+\mathcal{O}\left(\dfrac{\kappa_R^2}{v_1^2+4v_2^2}\right),\label{eq:gaugemass2}\\
    M_{Z''}^2&=g_B^2\left(v_1^2+4 v_2^2\right) \sec^2\epsilon+\dfrac{1}{4}\frac{g_R^2 g_Y^2 \left(\kappa _L^2+\kappa _R^2\right)}{ \left(g_R^2-g_Y^2\right)}\tan^2\epsilon.
    \label{eq:gaugemass3}
\end{align}
In the above equation,  we have traded $g_X$ in favor of $g_Y$ using the following coupling matching condition $1/g_Y^2 = 1/g_R^2 + 1/g_X^2$. 

\twocolumngrid
\bibliographystyle{utphys}
\bibliography{bib}
\end{document}